\documentclass[aps,preprint,showpacs,floatfix]{revtex4}
\usepackage{graphicx}

\begin{document}
\title{Superconducting nanobridges under magnetic fields}
\author{J.G. Rodrigo}
\author{H. Suderow}
\author{S. Vieira}
\thanks{Corresponding author: S. Vieira\\
e-mail: sebastian.vieira@uam.es\\
Fax: + 34 91 397 3961}
\affiliation{Laboratorio de Bajas Temperaturas, \\
Departamento de F\'isica de la Materia Condensada \\
Instituto de Ciencia de Materiales Nicol\'as Cabrera, Facultad de Ciencias \\
Universidad Aut\'onoma de Madrid, 28049 Madrid, Spain}
\date{\today}

\begin{abstract}
We report on the study of superconducting nanotips and nanobridges
of lead with a Scanning Tunnelling Microscope in tunnel and point
contact regimes. We deal with three different structures. A
nanotip that remains superconducting under a field of 2 Tesla. For
this case we present model calculations of the order parameter,
which are in good agreement with the experiments. An asymmetric
nanobridge of lead showing a two steps loss of the Andreev excess
current due to different heating and dissipation phenomena in each
side of the structure. A study of the effect of the thermal
fluctuations on the Josephson coupling between the two sides of a
superconducting nanobridge submitted to magnetic fields. The
different experiments were made under magnetic fields up to twenty
five times the volume critical field of lead, and in a temperature
range between 0.6 K and 7.2 K.
\end{abstract}

\maketitle

%%\flushbottom

\newpage

\section{Introduction}

Twenty years after the invention by G. Binnig and H. Rohrer of the
Scanning Tunneling Microscope (STM)\cite{Binnig}, the scientific
community has found a large diversity of interesting applications
to this instrument. In addition to its capability of obtaining
topographical information about conducting surfaces, the STM can
also be used to make local vacuum tunneling spectroscopy
measurements. It is frequent to find the acronym STM/S which
indicates an experimental set-up that is optimized to accomplish
both, microscopic and spectroscopic functions\cite{ReviewSTM}.

One of the most remarkable applications is the manipulation of
atoms. The capability to have a finger in the atomic realm has
permitted to a selected group of skilled scientists to build atom
by atom well defined structures on well characterized conducting
surfaces\cite{Eigler}. A fascinating world of new physical
phenomena and technological developments is born, as envisioned by
the gifted mind of R.P. Feynman already in 1959\cite{Feynman}.

The unprecedented control on the approach between macroscopic size objects
that the STM has introduced, has made possible experiments going from the
jump to the one atom contact, towards the evolution of connective necks of
very small sizes \cite{Gimzewski,Agrait93,Agrait93b}. The observed
conductance steps and its relationship to the quantization of the
conductance has triggered a considerable number of experimental and
theoretical works. In some superconducting elements, as Pb, Nb or Al,
measured at liquid helium temperatures, it has been possible, using the
highly non linear form of the subgap conductance due to Andreev reflection
processes, to relate the electronic conduction through a one atom contact to
the number of transmitting channels\cite{Scheer98}. This number turns out to
be characteristic of the electronic structure of the connecting atom, in
agreement with previous experiments and calculations for several normal
metals \cite{Sirvent96}.

A breakthrough in the world of nano-engineering has been the construction of
chains of atoms of gold and the characterization of its mechanical and
electrical properties\cite{Yanson98,Rubio01}. The forces involved in the
process of formation and the dissipation mechanisms have been investigated
in detail in this unique one-dimensional system. Recently, atomic chains of
gold atoms connecting superconducting electrodes were also created an studied%
\cite{Rubio02}. In this case, the electrodes were superconducting
lead, on top of which a thin film (of thickness 20nm) of gold was
evaporated, where superconductivity was induced by the proximity
effect.

In this paper we are going to focus on a superconducting
nanostructure that has been experimentally studied in our group
for the first time, and which consists of a superconducting
nanobridge connected to two normal bulky
electrodes\cite{JGR94a,Untiedt97,Poza98,Suderow00,Suderow00b,Suderow02}.
As we will show later on, the work of Prof. J.T. Devreese, to
which this volume is devoted, and his group in Antwerp have
contributed significantly to the understanding of its physical
behavior.

\section{Superconducting lead nanobridges}

\subsection{Fabrication and general behavior under magnetic fields}

Approaching carefully the tip and sample of an STM by measuring the
tunneling current for an applied bias voltage, it is possible to observe a
clear jump in the conductance when the mechanical contact between tip and
sample is attained. Pushing further on the tip into the sample, a neck of
variable cross-section is formed. The area of the minimal cross-section can
be estimated if the electronic mean free path is larger than the typical
dimensions of the contact. In this case, transport is ballistic and
Sharvin's formula applies. The neck can subsequently be elongated using the
displacement capabilities of the piezo element connected to the tip,
creating a nanobridge (NB) between both electrodes.

We have shown in previous papers that it is possible to induce order in the
bridge by means of a mechanical annealing procedure\cite{Agrait93b,Untiedt97}%
. A sequence of pushing and pulling processes of small amplitude are
superposed to a large, continuous pulling that defines the length of the
nanobridge.

During the formation of the NB bias voltage is kept constant, and the
current follows a staircase pattern that can be measured neatly during the
small amplitude pushing and pulling cycles. Often, very reproducible
patterns are measured, indicating that the neck evolves through very similar
atomic configurations. Simultaneous measurements of the conductance and the
force have shown that this staircase behavior of the conductance is due to a
sequence of plastic and elastic deformations occurring within the neck\cite
{AgraitPRL95}.

Simple models have been proposed to obtain good information about
the shape of the NB\cite{Untiedt97,chinos}. As it is known, lead
is a very ductile material which permits to obtain easily
nanobridges as long as several hundreds nanometers. The
nanostructure resulting from the fabrication process, which can be
followed until the rupture of the neck, consists of two similar
opposed cone-like sides. The NB can separated into two parts, and
we obtain two tips of nanoscopic dimensions with an apex of atomic
size. Due to the reduced atomic mobility at very low temperatures
the nanostructures do not change its form with time. Figure 1
shows schematically the above mentioned stuctures, as well as a
typical conductance curve obtained in tunnelling regime at 1.5 K
and zero magnetic field.

We are going to address now our attention to the study of these
structures when an external magnetic field is applied. It is well
established on thermodynamical grounds that to destroy
superconductivity with a magnetic field, the provided magnetic
energy must equal at least the condensation energy, which is given
by $\frac 12\mu _0H_c^2$, with $H_c$ the thermodynamic critical
field. Bulk lead is a type I superconductor whose critical field
is (at 0K) of about 800G. The characteristic superconducting
coherence and penetration lengths are $\xi _0=51$ nm and
$\lambda=32$ nm (also at 0K and H=0). Typical lead NB's have
lengths several times larger than $\xi _0$, but its lateral
dimensions are well below $\lambda $. In that case, there is no
significant Meissner screening, and therefore the magnetic field
needed to destroy superconductivity has to be much higher than the
bulk critical field, $H_{c1}$, in order to provide the required
magnetic energy. In this way, a superconducting nanobridge (SNB)
connected in a unique and perfect way to normal electrodes of the
same material is obtained, when the applied magnetic field exceeds
$H_{c1}$. The transport properties of these type of SNB have been
studied in detail, changing the diameter of the minimal cross
section, the temperature and the magnetic field.

Here we present recent results in SNB's created under high magnetic fields.
The process of formation and elongation of the nanobridge is continuously
controlled by a monitorization of the I-V curves at each elongation cycle.
When superconductivity nucleates in the nanobridge, characteristic
superconducting features appear in the I-V curves, determining the actual
creation of a SNB. We have measured the tunnelling density of states, after
completely breaking a SNB created under a magnetic field of 2T, and we
compare the results with some theoretical models. We also discuss two
interesting experiments, where we focus on the very different
superconducting behaviors found in asymmetric NB's due to the heating
induced by the circulating current, and we demonstrate how thermal
fluctuations influence superconductivity at nanoscopic dimensions.

\subsection{Superconducting lead nanotips (SNT) at 2T: experiments
and model calculations}

As a result of the rupture of a nanobridge, two, in principle equal,
nanotips were formed. The one corresponding to the STM tip is moved to
another position on the surface of the sample. The tunneling characteristics
of this N-S junction give us directly the superconducting density of states
of the nanotip, convoluted with the voltage derivative of the Fermi
function. In Fig.2(a) we represent the evolution (at 2 Tesla) of the
conductance curves of such a junction as a function of the temperature.
Superconducting features disappear near 4.5K, indicating that the tip
becomes normal.

These results can be understood quantitatively in the framework of the
Ginzburg Landau calculations of Misko, Fomin and Devreesse\cite{Misko01}.
These authors obtained in detail the spatial changes of the square of the
order parameter when a magnetic field higher than H$_{c1}$ is applied, using
a realistic geometry. In this way, they were able to map the amount of
superconducting phase along the SNB. Their calculations have made clear how
strong gradients of the superfluid concentration can be established at
magnetic fields much higher than the bulk critical field. In this way, these
authors significantly improved our first approach to the problem in which
the nanobridge was modelled by a wire of a few nanometers in diameter\cite
{Poza98}.

Following the characterization in the tunneling regime, the SNT
was brought into contact, with the same applied magnetic field (2
Tesla), without changing its geometry. This contact has a
resistance of 400 $\Omega $, and assuming that the conduction is
in the ballistic limit and using Sharvin's formula, its can be
estimated to correspond to a diameter of 0.8 nm . Transport in
this ballistic contact reflects the contribution of Andreev
reflection processes, as it can be seen in Fig.2b, where we plot
the conductance curves obtained at several temperatures between
0.8 and 4 K

We have analyzed these experiments following the model of \cite
{Suderow00}. The magnetic field enters as an effective, position
dependent pair breaking rate. The equations are solved
self-consistently, allowing to obtain a complete description of
the superconducting behaviour of the SNT in terms of energy and
distance to the tip apex. An important result of this theoretical
approach is that, in agreement with experiments, a gapless regime
is attained in the apex region already at a modest magnetic field.
The proximity effect from the neighboring normal regions plays a
crucial role to produce this singular situation, as it induces a
finite density of states at the Fermi level at the tip apex. The
SNT is represented as a conelike structure (inset of Fig.4) and
the fitting parameters used in the
calculation were $L_{cone}$=200 nm, $\xi $=25 nm, H=2 T, opening angle =15$%
^{\circ }$. The results of the calculation are represented in
Fig.3a. We want to remark that once the parameters are chosen in
order to fit the curve at the lowest temperature, they are kept
constant along the subsequent temperature variations. The
qualitative agreement is excellent. It is noteworthy to remark
that our experiment can be also modelled using a BCS density of
states with a pair breaking parameter $\Gamma ,$ as introduced in
Ref.\cite{Dynes}. Fitting the curve at the lowest temperature we
get $\Gamma =0.5\,$meV. This value is kept fixed in the whole
temperature range (as $H$
is also kept constant), so we can infer the temperature dependence of $%
\Delta $. In Fig.3b we plot the conductance vs bias curves calculated within
this analysis, and in Fig.4 we plot the temperature dependence of $\Delta $
from both models. Note that the result in Fig.4 is basically the same within
both approaches. However, there are tiny but relevant differences in the
calculation of the conductance. The model of Ref.\cite{Suderow00} gives much
better account of the V-shaped characteristics observed in the experiments
at low temperatures.

Note that the temperature dependence of $\Delta $ can also be
obtained using the curves in the contact regime shown in Fig.3b.
However, heating effects, which we treat below in more detail make
a precise interpretation difficult.

\subsection{Asymmetric superconducting bridges under magnetic fields}

As a result of the process of creation a of a SNB, we can find also
situations in which each half of the SNB presents a different geometry. The
asymmetry produces specific features in the I-V curves measured in contact
regime under magnetic field. An ideal curve in this case, without heating
effects, would present Josephson current at zero bias, the associated
subharmonic gap structure (related to $\Delta _1$and $\Delta _2$), and
finally a constant excess current equal to $\frac 8{3eR_N}\frac{\Delta
_1+\Delta _2}2$. A real curve, corresponding to an actual asymmetric SNB,
will present complex features due to the heating effects resulting of a
different power dissipation rate in each half. Thus, the expected loss of
the excess current will take place in a two step process. One part of the
SNB will become normal before the other as voltage and current (and
therefore, power) are increased along the acquisition of the I-V curve. It
is, therefore, the presence of heating the only way to unambiguously
determine the existence of an asymmetric SNB.

The actually measured I-V curve at a high external magnetic field
is somewhat more complex because it is the result of a competition
between the two consequences of having a SNB structure. Sharpening
the SNB produces an enhancement of superconductivity, but it also
leads to an increase of the heating effects on the I-V curve due
to a worse heat dissipation\cite{calent}. In fig.5 we present
curves corresponding to two different SNB geometries obtained in a
field of 0.6 Tesla, with the same normal state resistance. These
results illustrate the above mentioned situation. In fig. 5a,
curve A presents the highest value of Josephson current, thus it
corresponds to the largest value for $\Delta $ in both sides of
the neck, and therefore to the neck with the sharpest geometry.
This SNB presents the largest heating effects, as observed in the
two-step loss of the excess current which disappears completely at
4.8 mV (fig. 5b). Curve B, with a lower Josephson current,
corresponds to a less sharp SNB, with a better heat dissipation
rate. Correspondingly, the observed loss of the excess current
occurs at a higher voltage (7.3 mV, see fig. 5b) than in curve A.
The magnetic field behavior of these asymmetric bridges is shown
in Fig.6. While the position of the feature corresponding to the
first loss of the excess current (one of the sides of the SNB
becoming normal) remains almost constant, we observe that the
voltage at which the second loss takes place varies strongly with
the applied field.

Note that this behavior is only observed under magnetic fields, where the
different geometry of the two parts of the nanobridge gives two different
spatial evolutions of the superconducting order parameter. Obviously the
situation is more complicated, because the effective superconducting part of
the nanostructure is also changing when the current increases during the
heating process.

Clearly, it is possible to obtain relevant information about this
nanostructure by a careful analysis of the observed characteristic
curves. The capability of changing in-situ magnetic field,
cross-section, temperature and bias voltage gives the possibility
to study interesting non-equilibrium effects first reported in
\cite{Suderow00b}.

\subsection{Josephson coupling and thermal fluctuations in superconducting nanobridges under magnetic fields}

We discuss now another aspect of our measurements, related to the
Josephson effect in these nanobriges at $H>H_{c1}$. A zero field
study of these structures was previously reported in
ref.\cite{JJH0}. In superconducting bridges of resistances below a
thousand $\Omega $, the Josephson coupling energy is much higher
than the thermal energy in the relevant temperature range, and the
Josephson feature at zero bias is very clearly measured. When the
temperature is increased up to T$_c$ thermal fluctuations appear
in a very narrow temperature interval close to T$_c$. This leads
to the appearance of a finite conductance at zero bias when we
approache T$_c$. In Fig.7 we show the zero bias conductance as a
function of the reduced temperature for different magnetic fields.
At each magnetic field, we have identified the corresponding T$_c$
as the temperature where we loose superconducting features in the
I-V curves. At low magnetic fields, the conductance rapidly
increases when we cool below T$_c$. But at higher magnetic fields
(see Fig.7), when superconductivity becomes confined to a very
small region around the neck, we can observe a finite zero bias
conductance in a very large temperature range. This indicates that
the magnetic field dramatically increases thermal fluctuations
within the bridge. The change of size of the superconducting
region, which has shrunk from the bulk value to a nanometric size
superconducting bubble located at the center of the nanobridge,
must be on the origin of this effect. More detailed experiments
will be done changing the minimal cross section, the magnetic
field, temperature and smallest cross section in well
characterized bridges.

\section{Summary and conclusions}

In this paper, we have shown a characterization of several lead
nanostructures fabricated with the STM under magnetic fields.
Experimental results and model calculations of the density of
states of superconducting nanotips in the tunnel and point contact
regimes have been presented. We have also discussed a series of
features associated to asymmetric superconducting nanobridges
under magnetic field. New and interesting effects associated to a
non uniform heat dissipation in these structures have been
observed. The Josephson effect in nanometric size weak links under
magnetic fields and the role of thermal fluctuations was studied
by measuring one of these structures as a function of temperature.
The relevance of thermal fluctuations increases under magnetic
fields.

In conclusion, we remark the good agreement between experiments
and model calculations in these nanostructures. We also stress the
relevance of this type of nanostructures for the future
development of a new technology using elements of nanoscopic
dimensions.

It is a great pleasure for the authors to contribute to this volume in
homage to Prof. J.T. Devreesse. The discussions with Prof. J.T. Devreesse
and coworkers have revealed to us many important aspects of superconducting
nanostructures, a field where his contribution is of great relevance.

\section{Acknowledgements}

We also acknowledge discussions and helpful support from E.
Bascones, W. Belzig and F. Guinea. Financial support from the ESF
programme VORTEX and from the MCyT (Spain; grant
MAT-2001-1281-C02-0) is also acknowledged. The Laboratorio de
Bajas Temperaturas is associated to the ICMM of the CSIC.

\begin{figure}[ht]
\includegraphics[width=0.8\linewidth]{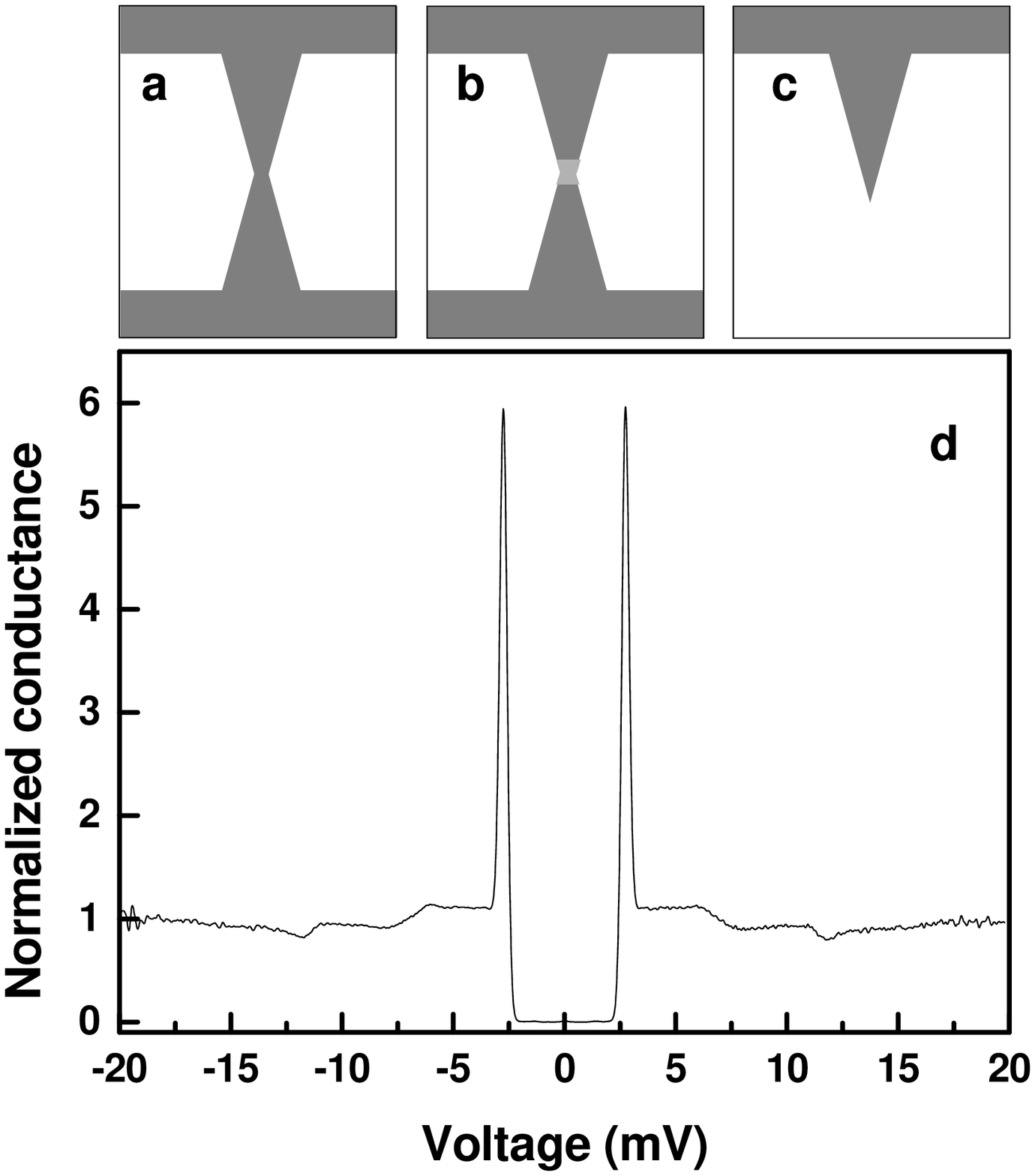}
\caption{The three main structures described in the text are shown
schematically: a) nanobridge (NB), b) neck, and c) nanotip (NT). In d) we
show a typical experimental conductance curve obtained in tunneling regime,
in zero magnetic field, at 1.5 K. (tunnel resistance is R$_N$=1.5 M$\Omega$%
). }
\label{fig:Fig1}
\end{figure}

\begin{figure}[ht]
\includegraphics[width=0.8\linewidth]{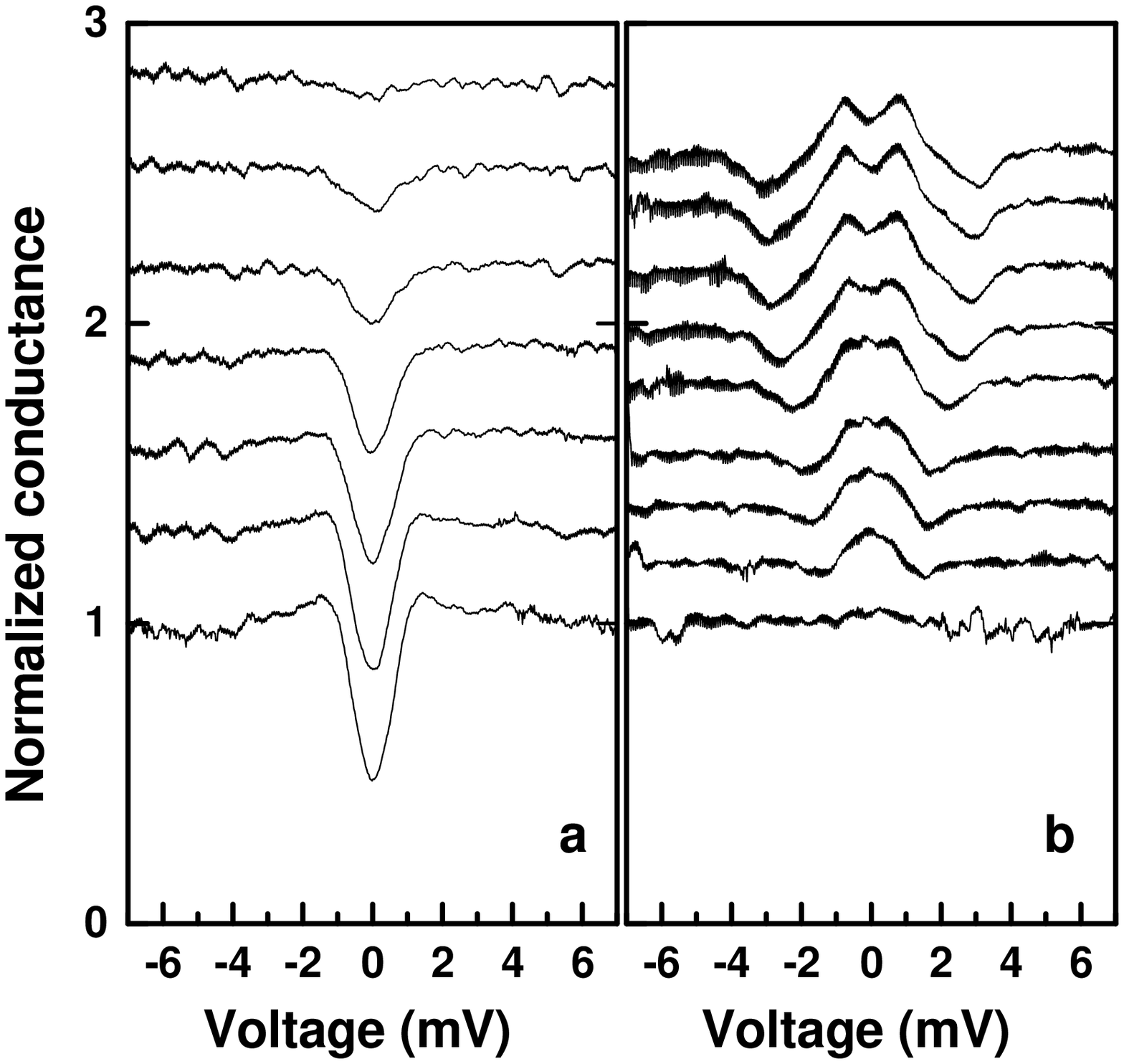}
\caption{(a.) Experimental conductance curves obtained in tunneling regime.
The tip was sharpened and elongated as described in the text. The external
magnetic field is 2 Tesla. Temperature is varied from 0.8 K to 4 K, from
bottom to top (curves are shifted for clarity). R$_N$=1.5 $M\Omega$. (b.)
Experimental conductance curves obtained in contact regime. Features due to
Andreev reflections and heating effects are clearly observable. The external
magnetic field is 2 Tesla. Temperature is varied from 0.8 K to 4 K, from top
to bottom (curves are shifted for clarity). R$_N$=400 $\Omega$. }
\label{fig:Fig2}
\end{figure}

\begin{figure}[ht]
\includegraphics[width=0.7\linewidth]{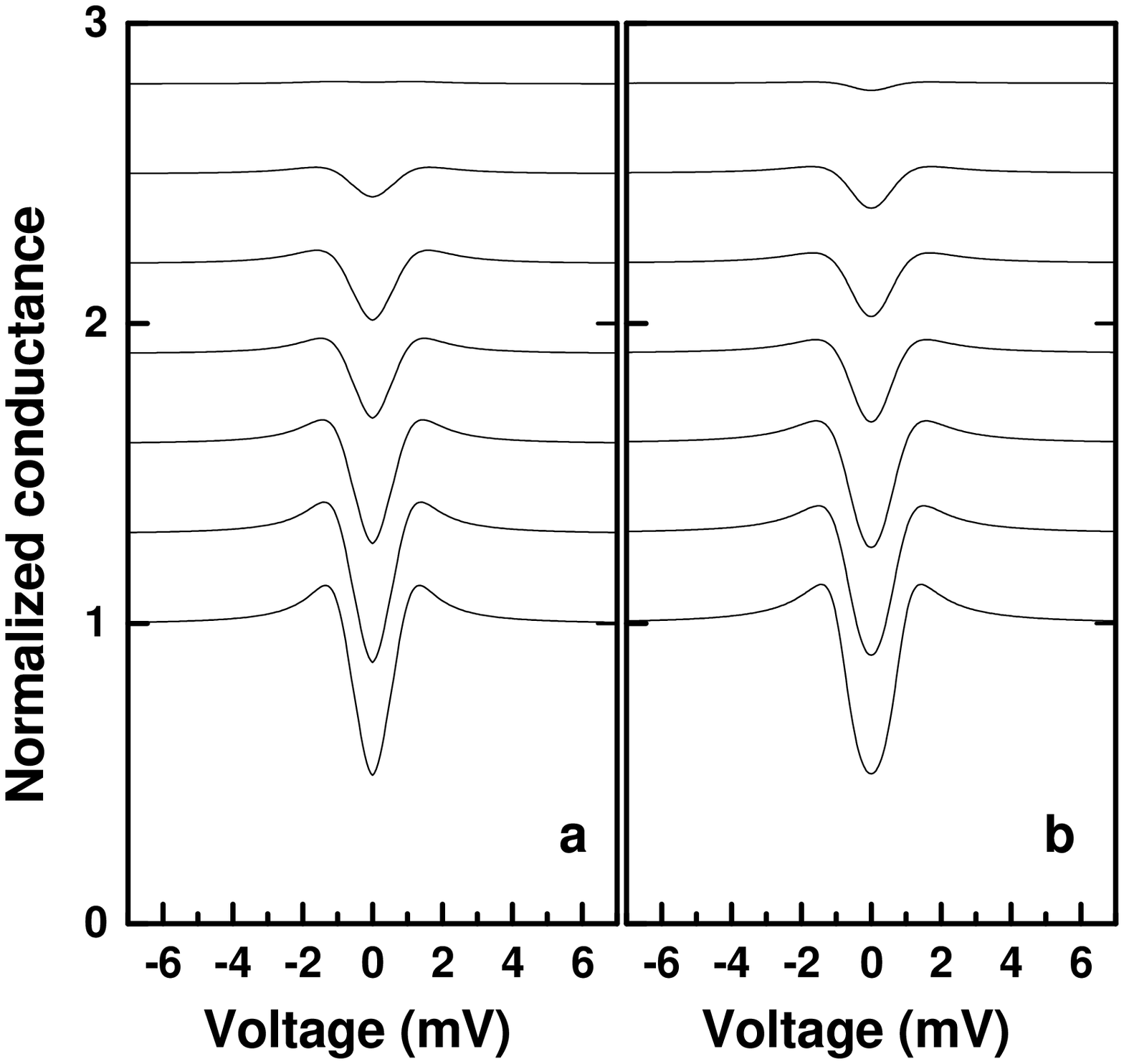}
\caption{ (a) Conductance curves obtained using the formalism of ref. \cite
{Suderow00} and the parameters indicated in the text. After fitting the
curve at the lowest temperature (bottom), all the parameters are kept
constant except temperature, that is varied from 1K to 4 K (in 0.5 K steps)
in order to reproduce the experimental results (curves are shifted for
clarity). (b) Conductance curves obtained after fitting the experimental
curves in tunneling regime using the Dynes formalism. The pair braking
parameter, $\Gamma$=0.5 meV, resulting after fitting the curve at the lowest
temperature (bottom), is kept constant for the rest of the curves. $\Delta$
is varied at each temperature (from 1K to 4 K, in 0.5 K steps) in order to
fit the experiment. (curves are shifted for clarity). }
\label{fig:Fig3}
\end{figure}

\begin{figure}[ht]
\includegraphics[width=0.7\linewidth]{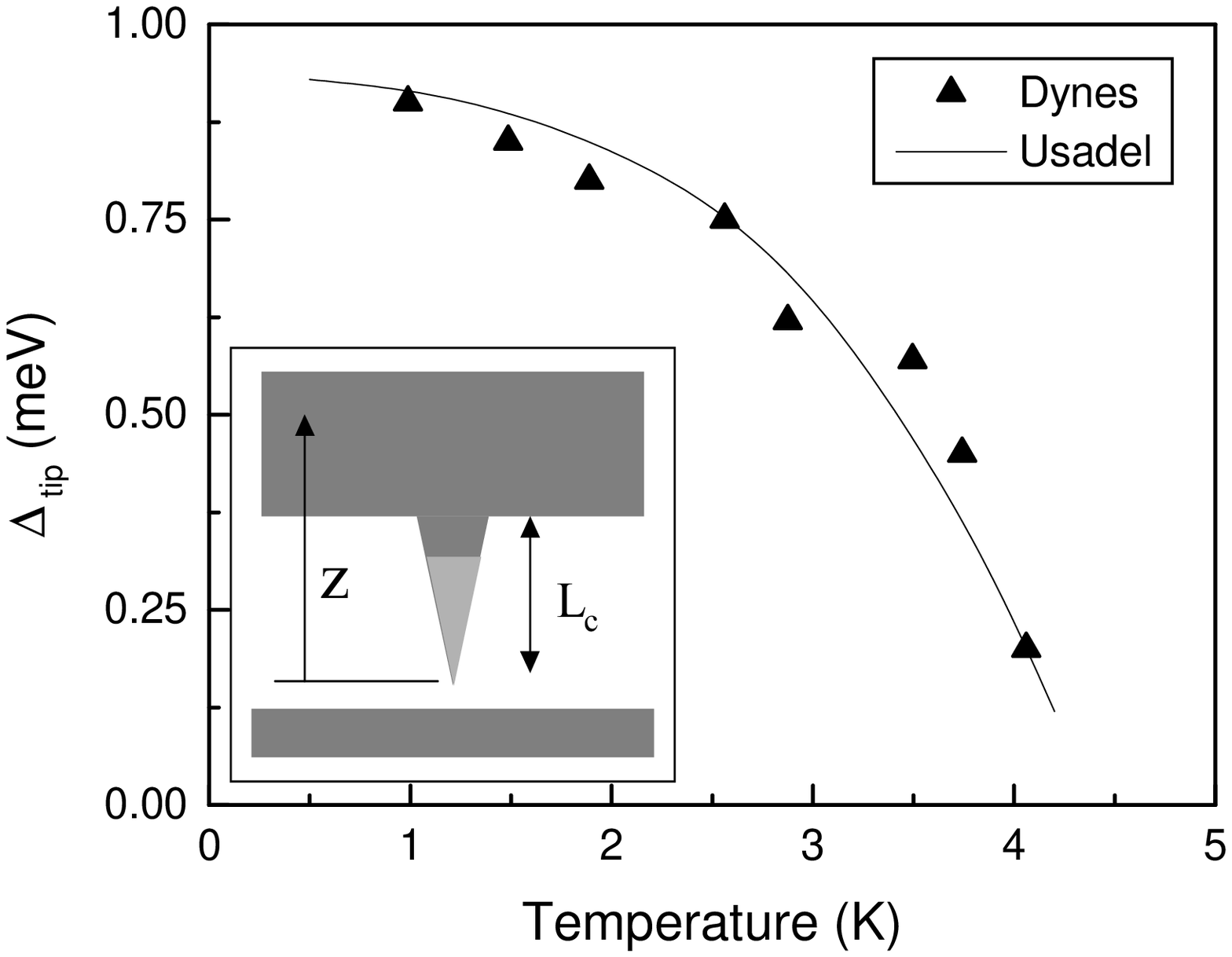}
\caption{Evolution of the superconducting gap at the tip apex vs temperature
obtained from the different models described in the text. Inset: Model
geometry used to reproduce the experimental evolution of the conductance
curves vs temperature. The tip is modelled as a long and sharp cone attached
to an infinite bulk electrode. The parameters used in the calculations were:
$L_{cone}$=200 nm, $\xi$=25 nm, H=2 T, opening angle =15$^{\circ}$. }
\label{fig:Fig4}
\end{figure}

\begin{figure}[ht]
\includegraphics[width=0.8\linewidth]{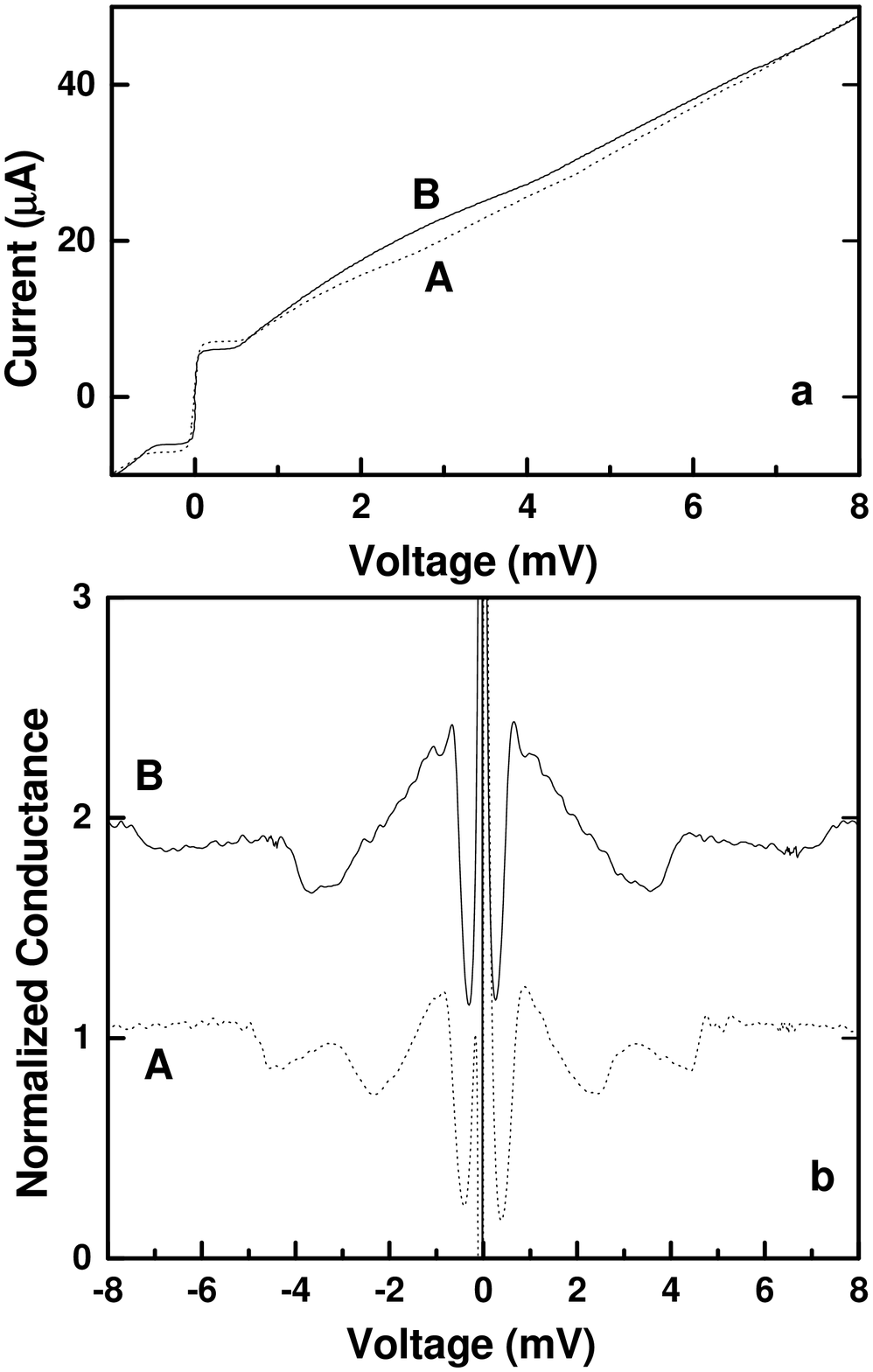}
\caption{I-V curves (a) and Conductance curves (b) obtained in contact
regime, in the presence of a magnetic field, H=0.6 Tesla, at 1 K,
corresponding to two different asymmetric nanobridges. The nanobridge with
the sharpest geometry (A, dotted line) presents an enhanced
superconductivity (larger Josephson current) and larger heating effects
(worse dissipation).In both cases R$_N$=180 $\Omega$.(Curves in (b) are
shifted for clarity)}
\label{fig:Fig5}
\end{figure}

\begin{figure}[ht]
\includegraphics[width=0.8\linewidth]{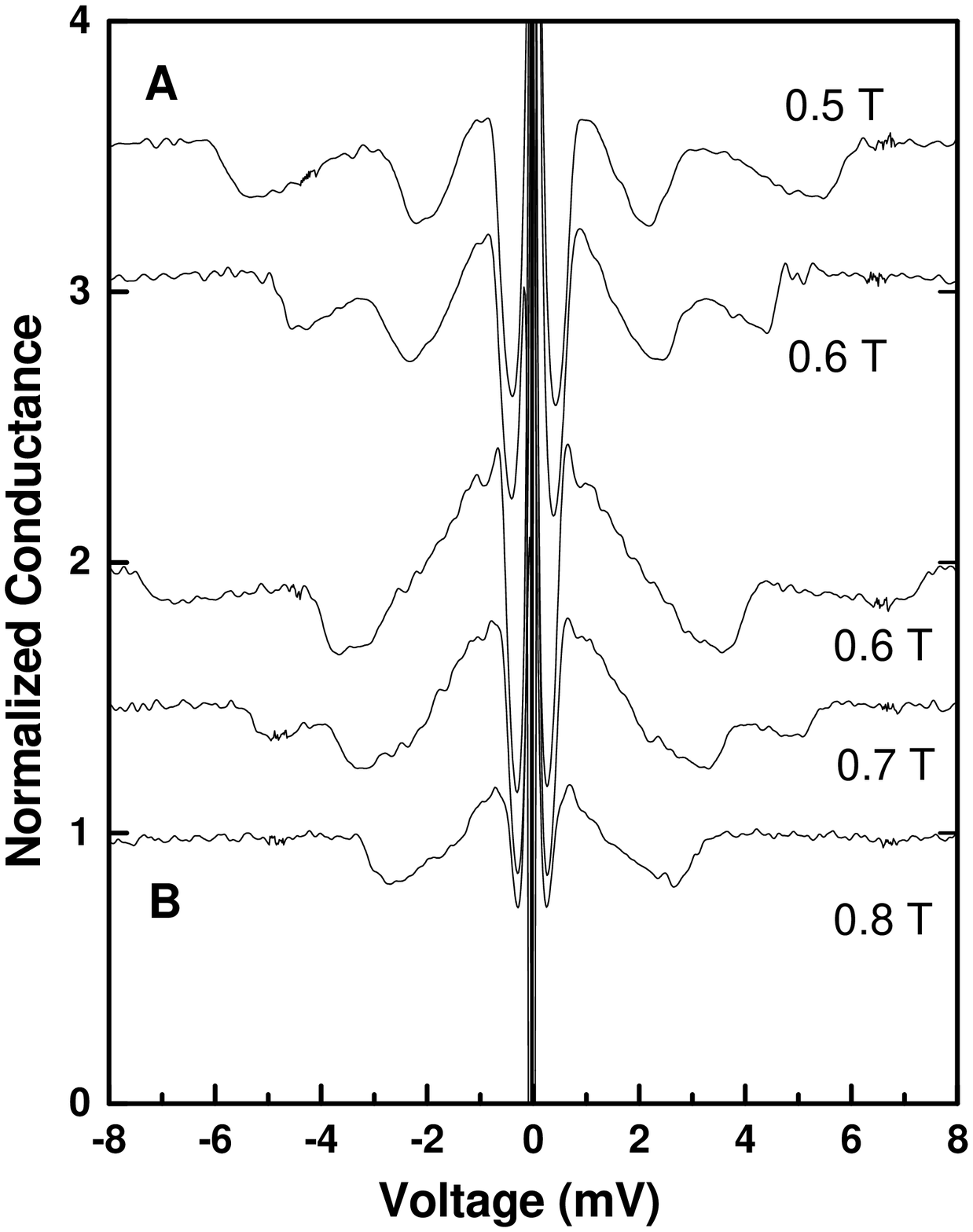}
\caption{Conductance curves obtained at different magnetic fields for the
two nanobridges presented in fig.5. In all cases R$_N$=180 $\Omega$ and T=1
K.(Curves are shifted for clarity)}
\label{fig:Fig6}
\end{figure}

\begin{figure}[ht]
\includegraphics[width=0.8\linewidth]{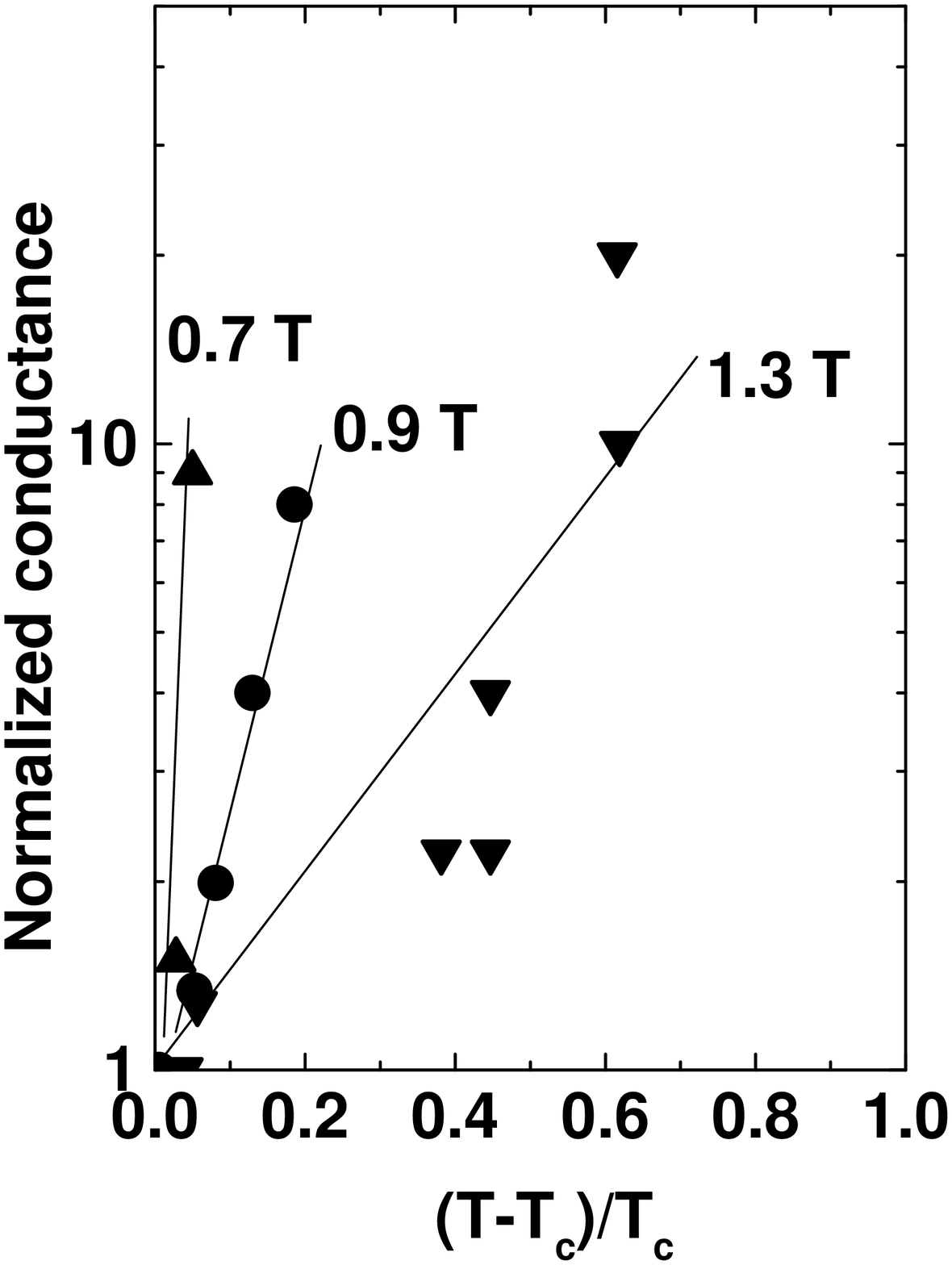}
\caption{Evolution of the zero bias conductance for a SNB as a
function of the reduced temperature $(T-T_c)/T_c$, for various
magnetic fields (T$_c$ is measured for each field, see text; lines
are guides to the eye). The smallest contact is of low resistance
(about 100 $\Omega$) and the SNB has a zero temperature critical
field of about 1.7 T. The temperature regime where phase
fluctuations appear considerably increases at high magnetic
fields.} \label{fig:Fig7}
\end{figure}


\begin{references}
\bibitem{Binnig}  G. Binnig, H. Rohrer, Ch. Gerber and E. Weibel, Appl.
Phys. Lett. 40, 178 (1982)

\bibitem{ReviewSTM}  See e.g. C.J. Chen, ''Introduction to Scanning
Tunneling Microscopy'', Oxford Series in Optical and Imaging Sciences,
Oxford University Press (1993).

\bibitem{Eigler}  M.F. Crommie, C.P. Lutz, D.M. Eigler. Science 262, 218-220 (1993).

\bibitem{Feynman}  ``There's Plenty of Room at the Bottom'' (1959) in
Richard P. Feynman, Jeffrey Robbins. The Pleasure of Finding Things Out: the
best short works. Penguin. 1999

\bibitem{Gimzewski}  J.K. Gimzewski and R. Moller, Phys. Rev. B {\bf 36},
1284 (1987).

\bibitem{Agrait93}  N. Agra{\"\i }t, J. G. Rodrigo and S. Vieira, Phys. Rev.
B {\bf 47}, 12345 (1993).

\bibitem{Agrait93b}  N. Agra{\"\i }t, J. G. Rodrigo, C. Sirvent and S.
Vieira, Phys. Rev. B {\bf 48}, 8499 (1993).

\bibitem{Scheer98}  E. Scheer, N. Agra{\"\i }t, A. Cuevas, A. Levy-Yeyati,
B. Ludolph, A. Mart{\'\i }n-Rodero, G. Rubio, J. M. van Ruitenberg and C.
Urbina, Nature {\bf 394}, 154 (1998).

\bibitem{Sirvent96}  C. Sirvent, J. G. Rodrigo, S. Vieira, L. Jurczyszyn, N.
Mingo, and F. Flores, Phys. Rev. B {\bf 53}, 16086-16090 (1996).

\bibitem{Yanson98}  Yanson et al., Nature {\bf 395}, 783 (1998).

\bibitem{Rubio01}  G. Rubio-Bollinger, S. R. Bahn, N. Agra\"\i t, K. W.
Jacobsen, and S. Vieira Phys. Rev. Lett. 87, 026101 (2001)

\bibitem{Rubio02}  G. Rubio-Bollinger et al., submitted.

\bibitem{JGR94a}  J.G. Rodrigo, N. Agra{\"\i }t and S. Vieira, Phys. Rev. B,
{\bf 50}, 374 (1994).

\bibitem{Untiedt97}  C. Untiedt, G. Rubio, S. Vieira and N. Agra{\"\i }t,
Phys. Rev. B {\bf 56}, 2154 (1997).

\bibitem{Poza98}  M. Poza, E. Bascones, J. G. Rodrigo, N. Agra{\"\i }t, S.
Vieira and F. Guinea, Phys. Rev. B {\bf 58}, 11173 (1998).

\bibitem{Suderow00}  H. Suderow, E. Bascones, W. Belzig, F. Guinea, S.
Vieira Europhysics Letters, {\bf 50}, 749 (2000).

\bibitem{Suderow00b}  H. Suderow, S. Vieira, Physics Letters A, {\bf 275},
299 (2000).

\bibitem{Suderow02}  H. Suderow, E. Bascones, A. Izquierdo, F. Guinea, S.
Vieira, Phys Rev. B, {\bf 65}, 100519 (R) (2002).

\bibitem{AgraitPRL95}  N. Agra\"\i t et al. Phys. Rev. Lett., {\bf 74}, 3995
(1995); G. Rubio et al. Phys. Rev. Lett. {\bf 76}, 2032 (1996); N. Agra\"\i
t et al. Thin Solid Films, {\bf 253}, 199 (1994).

\bibitem{chinos}  Z. Gai et al., Phys. Rev. B, {\bf 58}, 2185 (1998).

\bibitem{Misko01}  V.R. Misko, V.M. Fomin and J.T. Devreese, Phys. Rev. B,
{\bf 64}, 14517 (2001).

\bibitem{Dynes}  R. C. Dynes, V. Narayamurti and J. P. Garno, Phys.
Rev.Lett. {\bf 41}, 1509 (1978).

\bibitem{calent}  M. Tinkham et al. J. of Applied Physics, {\bf 48} , 1311
(1977).

\bibitem{JJH0}  J.G. Rodrigo, N. Agra{\"\i }t, C. Sirvent and S. Vieira, et
al. Phys. Rev. B {\bf 50}, 12788 (1994).

\bibitem{Fluct}  See e.g. A. Barone and G. Paterno, ``Physics and
Applications of the Josephson Effect'', Wiley, New York (1982) and K.K.
Likharev, ``Dynamics of Josephson junctions and circuits'', Gordon and
Breach, 1986.
\end{references}
\end{document}